\documentclass[preprint,pra,showpacs,nofootinbib]{revtex4}
\usepackage[mathscr]{euscript}
\usepackage[dvips]{pstcol}
\usepackage{graphicx}
\usepackage{float,epsfig}
\usepackage{dcolumn}
\usepackage{bm}
\usepackage{graphicx}
\usepackage{dcolumn}
\usepackage{amsmath,amssymb,amsthm}
\usepackage{subfigure}
\usepackage[colorlinks=true,linkcolor=blue]{hyperref}
\newcommand{\bea}{\begin{eqnarray}}
\newcommand{\eea}{\end{eqnarray}}
\newcommand{\beq}{\begin{equation}}
\newcommand{\eeq}{\end{equation}}

\def\/{\over}

\begin{document}

\title{The formalism for energy changing rate of an accelerated atom coupled with electromagnetic vacuum fluctuations}
\author{Anwei Zhang}
\email{hunnuzaw@163.com}
\affiliation{Institute of Physics and Key Laboratory of Low
Dimensional Quantum Structures and Quantum
Control of Ministry of Education,\\
Hunan Normal University, Changsha, Hunan 410081, China \\}


\begin{abstract}
The structure of the rate of variation of the atomic energy for an arbitrary stationary motion of the atom in interaction with a quantum electromagnetic field is investigated. Our main purpose is to rewrite the formalism in Ref. \cite{zz} and to deduce the general expressions of the Einstein $A$ coefficients of an atom on an arbitrary stationary trajectory. The total rate of change of the energy and Einstein coefficients of the atom near a plate with finite temperature or acceleration are also investigated.
\end{abstract}

\maketitle

\section{Introduction}
\label{intro}
Spontaneous emission is one of the most important
features of atoms and so far mechanisms such as vacuum
fluctuations \cite{Welton, Compagno}, radiation reaction \cite{Ackerhalt},  or a combination
of them \cite{Milonni} have been put forward to explain why spontaneous emission occurs. The ambiguity in physical interpretation
arises because of the freedom in the choices of ordering of commuting
operators of atom and field in a Heisenberg picture approach to the problem. The controversy was resloved, when
Dalibard, Dupont-Roc and Cohen-Tannoudji \cite{ddc1, ddc2} proposed a formalism which distinctively separates
the contributions of vacuum fluctuations and radiation reaction to the rate of change of an atomic observable
by demanding a symmetric operator ordering of atom and field variables.

Later, the formalism of Ref. \cite{ddc2} has been generalized by Audretsch, M\"{u}ller and Holzmann  to evaluate vacuum fluctuations and radiation reaction contributions to the spontaneous excitation rate \cite{audretsch} and radiative energy shifts
of an accelerated two-level atom interacting with a scalar field in a unbounded Minkowski space. Based upon this new
formalism, the radiative energy shifts and spontaneous excitation of a hydrogen
atom, moving with uniform acceleration and interacting with the electromagnetic field, are considered \cite{Passante,zz}.
 Their studies show that the effect of electromagnetic vacuum fluctuations on atom, contrarily to the scalar field case, contains a nonthermal
 acceleration-dependent correction and the contribution of radiation reaction is affected by the acceleration.

However, the physical meaning of the previous formalisms \cite{audretsch, zz} are not obvious in the form and the calculations are too burdensome \cite{yu}. It is then natural for us to wonder whether there exists other formalism which can express the physical meaning more directly and simplify the calculations. In this paper, by treating a two-level atom in a bath of fluctuating electromagnetic field in vacuum, we plan to address this issue by investigating the formalism in Ref. \cite{zz}, and rewriting the rate of variation of the mean atomic energy in a new meaningful form. Then we deduce the general form of the Einstein $A$ coefficients corresponding to two spontaneous processes. We will also study the rate of change of the atomic energy and Einstein coefficients in the cases that the atom is placed near a reflecting boundary at finite temperature or acceleration.

\section{Vacuum fluctuations and radiation reaction}
\label{sec:vacuum fluctuations}

We consider the interaction of a pointlike two-level atom on an arbitrary stationary trajectory $x(\tau)=\big(t(\tau), \vec{x}(\tau)\big)$ and the quantum electromagnetic field. $\tau$ denotes the proper time on the trajectory. The stationary trajectory, which follows the orbits of a timelike Killing vector field, guarantees the existence of stationary atomic states, the excited state $|+\rangle$ and the ground state $|-\rangle$, with energies $\pm\frac{1}{2}\omega_0$ and a level spacing $\omega_0$.
The Hamiltonian which controls the time evolution of the atom with respect to $\tau$ takes the form $H_A=\frac{1}{2}\omega_0 \sigma_3$, where
$\sigma_3=|+\rangle\langle+|-|-\rangle\langle-|$ is the Pauli matrix and we set $\hbar=c=k_B=1$ throughout the text.
The Hamiltonian that describes the interaction between the atom and the  electromagnetic field in the dipole coupling scheme can be written as $H_I(\tau)=-e\mathbf{r}\cdot \mathbf{E}(x(\tau))=-e\Sigma_{mn}\mathbf{r}_{mn}\cdot \mathbf{E}(x(\tau))\sigma_{mn}(\tau)$,
where $e$ is the electron electric charge, $e\mathbf{r}$ the atomic electric dipole moment, $\mathbf{E}(x)$ the electric field strength and
$\sigma_{mn}(\tau)=|m\rangle\langle n|$ with $m, n$ referring to $\{+, -\}$.

Now the Heisenberg equations of motion for the atom and field observables can be written down. We will isolate the two physical
mechanisms that contribute to the rate of variation of atomic observables: the contribution of vacuum fluctuations and that of radiation reaction.
For this purpose, we split the solutions of the equations of motion into free and source part. Following the forms of Ref. \cite{Passante,zz}, the contributions of vacuum fluctuations$(vf)$ and
radiation reaction $(rr)$ to the rate of change of the mean atomic excitation energy can be written as
\begin{eqnarray}\label{1}
  \bigg\langle\frac{dH_A(\tau)}{d\tau}\bigg\rangle_{vf} &=& 2ie^{2}\int^{\tau}_{\tau_0}d\tau'C^{F}_{ij}(x(\tau), x(\tau'))\frac{d}{d\tau}(\chi^{A}_{ij})_b(\tau, \tau'),
\end{eqnarray}
\begin{eqnarray}\label{2}
 \bigg\langle\frac{dH_A(\tau)}{d\tau}\bigg\rangle_{rr} &=& 2ie^{2}\int^{\tau}_{\tau_0}d\tau'\chi^{F}_{ij}(x(\tau), x(\tau'))\frac{d}{d\tau}(C^{A}_{ij})_b(\tau, \tau'),
\end{eqnarray}
with $|\rangle=|b, 0\rangle$ representing the atom in the initial state $|b\rangle$ and the field in the vacuum state $|0\rangle$. Note that repeated indices $i, j$ are summed in the paper.
The explicit forms of the statistical functions of the atom are given by
\begin{eqnarray}\label{3}
  (C^{A}_{ij})_b(\tau, \tau') = \frac{1}{2}\sum_d [\langle b|r_i(0)|d\rangle\langle d|r_j(0)|b\rangle e^{i\omega_{bd}(\tau-\tau')}
 +\langle b|r_j(0)|d\rangle\langle d|r_i(0)|b\rangle e^{-i\omega_{bd}(\tau-\tau')}],
  \end{eqnarray}
\begin{eqnarray}\label{4}
 (\chi^{A}_{ij})_b(\tau, \tau') = \frac{1}{2}\sum_d [\langle b|r_i(0)|d\rangle\langle d|r_j(0)|b\rangle e^{i\omega_{bd}(\tau-\tau')}
 -\langle b|r_j(0)|d\rangle\langle d|r_i(0)|b\rangle e^{-i\omega_{bd}(\tau-\tau')}],
  \end{eqnarray}
  where $\omega_{bd}=\omega_b-\omega_d$ and the sum extends over a complete set of atomic states.
The symmetric correlation function and linear susceptibility of the field are defined as
\begin{eqnarray}\label{5}
  C^{F}_{ij}(x(\tau),x(\tau')) &=& \frac{1}{2}\langle 0|\{E^{f}_i(x(\tau)), E^{f}_j(x(\tau'))\}|0\rangle,
 \end{eqnarray}
 \begin{eqnarray}\label{6}
  \chi^{F}_{ij}(x(\tau),x(\tau')) &=& \frac{1}{2}\langle 0|[E^{f}_i(x(\tau)), E^{f}_j(x(\tau'))]|0\rangle.
\end{eqnarray}

\section{Formalism}
\label{sec:Formalism}
Based on the formalism given above, in this section we will rewrite the form of the relaxation rates (\ref{1}), (\ref{2})
for atoms in arbitrary stationary motion which ensures that the correlation functions are homogeneous in time, since stationary motion has a
characterization that the geodesic distance between two points $x(\tau)$ and $x(\tau')$ on the trajectory depends only on the proper time interval $\tau-\tau'$ \cite{aud}. So we can define
\begin{equation}\label{19}
  \langle0|E^{f}_i(x(\tau))E^{f}_j(x(\tau'))|0\rangle\equiv G_{ij}(\tau-\tau').
\end{equation}
Now, inserting (\ref{4}), (\ref{5}) in (\ref{1}) and taking $u=\tau-\tau'$, we can get
\begin{eqnarray}
 \bigg\langle\frac{dH_A(\tau)}{d\tau}\bigg\rangle_{vf} =
 -\frac{e^{2}}{2}\sum_d\omega_{bd}\langle b|r_i(0)|d\rangle\langle d|r_j(0)|b\rangle\big[\mathcal{G}_{ij}(\omega_{bd})+\mathcal{G}_{ji}(-\omega_{bd})\big],
\end{eqnarray}
where $\mathcal{G}_{ij}(\lambda) = \int^{\infty}_{-\infty}du e^{i\lambda u}G_{ij}(u)$ is the Fourier transforms of the field correlation functions. For two-level atom, we can split
this expression into the case $\omega_b>\omega_d$,
\begin{equation}\label{aa}
\bigg\langle\frac{dH_A(\tau)}{d\tau}\bigg\rangle_{vf}=-\frac{e^{2}}{2}\omega_0\langle +|r_i(0)|-\rangle\langle -|r_j(0)|+\rangle(\mathcal{G}_{ij}(\omega_0)+\mathcal{G}_{ji}(-\omega_0)),
\end{equation}
and the case $\omega_b<\omega_d$,
\begin{equation}\label{bb}
\bigg\langle\frac{dH_A(\tau)}{d\tau}\bigg\rangle_{vf}=\frac{e^{2}}{2}\omega_0\langle -|r_i(0)|+\rangle\langle +|r_j(0)|-\rangle(\mathcal{G}_{ij}(-\omega_0)+\mathcal{G}_{ji}(\omega_0)).
\end{equation}
If we define
\begin{eqnarray}\label{zz}
  \mathcal{G}^{+}(\omega_0)&=&e^{2}\langle +|r_i(0)|-\rangle\langle -|r_j(0)|+\rangle\mathcal{G}_{ij}(\omega_0)\nonumber\\
  &=&e^{2}\langle +|r_i(0)|-\rangle\langle -|r_j(0)|+\rangle \int^{\infty}_{-\infty} d u e^{i\omega_0 u}\langle0|E^{f}_i(x(\tau))E^{f}_j(x(\tau'))|0\rangle
 \end{eqnarray}
 and
 \begin{eqnarray}\label{zzz}
 \mathcal{G}^{-}(-\omega_0)&=&e^{2}\langle -|r_i(0)|+\rangle\langle +|r_j(0)|-\rangle\mathcal{G}_{ij}(-\omega_0)\nonumber\\
 &=&e^{2}\langle -|r_i(0)|+\rangle\langle +|r_j(0)|-\rangle \int^{\infty}_{-\infty} d u e^{-i\omega_0 u}\langle0|E^{f}_i(x(\tau))E^{f}_j(x(\tau'))|0\rangle,
\end{eqnarray}
(\ref{aa}) and (\ref{bb}) can be written as
\begin{eqnarray}\label{s}
 \bigg\langle\frac{dH_A(\tau)}{d\tau}\bigg\rangle_{vf}=\Bigg\{
         \begin{array}{r}
 -\frac{1}{2}\omega_0(\mathcal{G}^{+}(\omega_0)+ \mathcal{G}^{-}(-\omega_0)), \;\;for \;initial \;excited\; state,\\
   \frac{1}{2}\omega_0(\mathcal{G}^{+}(\omega_0)+\mathcal{G}^{-}(-\omega_0)), \;\;for \;initial \;ground\; state.\\
          \end{array}
\end{eqnarray}

In a similar way, we can obtain
\begin{eqnarray}\label{9}
 \bigg\langle\frac{dH_A(\tau)}{d\tau}\bigg\rangle_{rr}=\Bigg\{
         \begin{array}{r}
 -\frac{1}{2}\omega_0(\mathcal{G}^{+}(\omega_0)- \mathcal{G}^{-}(-\omega_0)), \; \;excited\; state,\\
   -\frac{1}{2}\omega_0(\mathcal{G}^{+}(\omega_0)-\mathcal{G}^{-}(-\omega_0)), \;\;ground\; state.\\
          \end{array}
\end{eqnarray}

Finally, we add the contributions of vacuum fluctuations (\ref{s}) and radiation reaction (\ref{9}) to obtain the total rate of change of the atomic energy
\begin{eqnarray}\label{10}
 \bigg\langle\frac{dH_A(\tau)}{d\tau}\bigg\rangle_{tot}=\Bigg\{
         \begin{array}{r}
 -\omega_0\mathcal{G}^{+}(\omega_0), \; \;excited\; state,\\
   \omega_0\mathcal{G}^{-}(-\omega_0), \;\;ground\; state.\\
          \end{array}
\end{eqnarray}
As can be seen, in the excited state, the rate of variation of atomic energy is the quantity $-\omega_0\mathcal{G}^{+}(\omega_0)$, and in the ground state, it is $\omega_0\mathcal{G}^{-}(-\omega_0)$. So $\mathcal{G}^{+}(\omega_0)$ has the physical meaning of emission rate, $\mathcal{G}^{-}(-\omega_0)$ the meaning of excitation rate.
\section{Einstein $A$ coefficients}
\label{sec:Einstein Coefficients}
Next we will investigate the exact physical meaning of $\mathcal{G}^{+}(\omega_0)$ and $\mathcal{G}^{-}(-\omega_0)$. On the
condition that atom moves on arbitrary stationary trajectory, we have two spontaneous processes, i.e., the spontaneous excitation and de-excitation. Thus there are two Einstein coefficients $A_\uparrow$ and $A_\downarrow$ which
describe the corresponding transition rates.
Consider an ensemble of $N$ atoms. Let $N_1$ denote the number of atoms
in the ground state, $N_2$ the number in the excited state. The
rate equations are given by
\begin{equation}\label{11}
  \frac{dN_2}{d\tau}=-\frac{dN_1}{d\tau}= N_1A_\uparrow- N_2A_\downarrow
\end{equation}
with
\begin{equation}
\langle H_A\rangle=\frac{1}{N}\bigg(-\frac{1}{2}\omega_0N_1+\frac{1}{2}\omega_0N_2\bigg).
\end{equation}
The solution of the above equations is
\begin{equation}\label{12}
  \langle H_A(\tau)\rangle=-\frac{1}{2}\omega_0+\frac{A_\uparrow}{A_\uparrow+A_\downarrow}\omega_0+\bigg(\langle H_A(0)\rangle+\frac{1}{2}\omega_0-\frac{A_\uparrow}{A_\uparrow+A_\downarrow}\omega_0\bigg)e^{-(A_\uparrow+A_\downarrow)\tau}.
\end{equation}
On the other hand, we can simplify (\ref{10}) by using \cite{audretsch}
\begin{eqnarray}\label{qq}
  \sum_{\omega_a<\omega_b}\omega^{2}_{ab}|\langle a|\sigma^{f}_2(0)|b\rangle|^{2}\pm \sum_{\omega_a>\omega_b}\omega^{2}_{ab}|\langle a|\sigma^{f}_2(0)|b\rangle|^{2}=\Bigg\{
         \begin{array}{r}
  \omega^{2}_0          \\
   -2\omega^{2}_0\langle a|\frac{1}{2}\sigma^{f}_3(0)|a\rangle.\\
          \end{array}
\end{eqnarray}
In order $\mu^{2}$, $\omega_0\langle\frac{1}{2}\sigma^{f}_3(0)\rangle$ can be replaced by $\langle H_A\rangle$.
Substituting (\ref{qq}) in (\ref{10}) which is rewritten by $\sum_{\omega_a<\omega_b}$ and $\sum_{\omega_a>\omega_b}$, we obtain a differential equation for $\langle H_A\rangle$
\begin{equation}\label{13}
 \bigg\langle\frac{dH_A(\tau)}{d\tau}\bigg\rangle=\frac{1}{2}\omega_0(\mathcal{G}^{-}(-\omega_{0})
 -\mathcal{G}^{+}(\omega_{0}))-(\mathcal{G}^{-}(-\omega_{0})+\mathcal{G}^{+}(\omega_{0}))\langle H_A(\tau)\rangle,
\end{equation}
the solution of which gives the time evolution of the mean
atomic energy
\begin{eqnarray}\label{14}
  \langle H_A(\tau)\rangle &=&-\frac{1}{2}\omega_0+\frac{\mathcal{G}^{-}(-\omega_{0})}{\mathcal{G}^{-}(-\omega_{0})+\mathcal{G}^{+}(\omega_{0})}\omega_0
  +\bigg(\langle H_A(0)\rangle\nonumber \\&&+\frac{1}{2}\omega_0-\frac{\mathcal{G}^{-}(-\omega_{0})}{\mathcal{G}^{-}(-\omega_{0})
  +\mathcal{G}^{+}(\omega_{0})}\omega_0\bigg)e^{-(\mathcal{G}^{-}(-\omega_{0})+\mathcal{G}^{+}(\omega_{0}))\tau}.
\end{eqnarray}
Comparing (\ref{12}) with (\ref{14}), we can easily identify the Einstein coefficients $A_\downarrow$ and $A_\uparrow$  as the Fourier transforms of two-point correlation function for electric fields
\begin{eqnarray}\label{29}
A_\downarrow&=&e^{2}\langle +|r_i(0)|-\rangle\langle -|r_j(0)|+\rangle \int^{\infty}_{-\infty} d u e^{i\omega_0u}
\langle0|E^{f}_i(x(\tau))E^{f}_j(x(\tau'))|0\rangle,\nonumber\\
A_\uparrow&=&e^{2}\langle -|r_i(0)|+\rangle\langle +|r_j(0)|-\rangle \int^{\infty}_{-\infty} d u e^{-i\omega_0u}
\langle0|E^{f}_i(x(\tau))E^{f}_j(x(\tau'))|0\rangle,
\end{eqnarray}
with $u=\tau-\tau'$.
These two equations are the general forms for the Einstein spontaneous emission coefficients and spontaneous excitation coefficients of atom on arbitrary stationary trajectory.  Once we work out the explicit $\mathcal{G}^{+}(\omega_0)$, $\mathcal{G}^{-}(-\omega_0)$, we can directly give out Einstein coefficients and $\bigg\langle\frac{dH_A(\tau)}{d\tau}\bigg\rangle_{tot}$. Thus the calculations will be greatly simplified. In other word, the present work generalizes and simplifies previous works.

\section{The case of an atom near a reflecting boundary at finite temperature}
\label{sec:example1}
In what follows, we will apply the previously developed formalism to study the rate of change of the energy of an atom immersed in a thermal bath of external field with a boundary at $z=0$.
In such a thermal case, the vacuum expectation value in~(\ref{5}), (\ref{6}), (\ref{19}) and then (\ref{zz}), (\ref{zzz}) should be
replaced by thermal average \cite{audretsch}. The two point function of the electric field four potential $A^{\mu}$ can be written as \cite{jhep}
\begin{eqnarray}\label{a1}
  \langle0|A^{\mu}(x)A^{\nu}(x')|0\rangle_{\beta}&=&
  \frac{1}{4\pi^{2}}\sum^{\infty}_{k=-\infty}\bigg[\frac{\eta^{\mu\nu}}{(t-t'+ik\beta-i\varepsilon)^{2}-(x-x')^{2}-(y-y')^{2}-(z-z')^{2}}\nonumber\\
  &&-\frac{\eta^{\mu\nu}+2n^{\mu}n^{\nu}}{(t-t'+ik\beta-i\varepsilon)^{2}-(x-x')^{2}-(y-y')^{2}-(z+z')^{2}}\bigg],
\end{eqnarray}
where $\beta$ is the reverse temperature of the thermal bath, $\varepsilon\rightarrow+0,$  $\eta^{\mu\nu}=diag(1,-1,-1,-1)$ and $n^{\mu}=(0,0,0,1)$ is the unit normal vector to the boundary.
The first term of right side of~(\ref{a1}) refers to the two point function in free space and the last term gives the correction due to the boundary.
The electric field two point function can then be obtained as follows
\begin{equation}\label{a2}
\langle0|E^{f}_i(x(\tau))E^{f}_j(x(\tau'))|0\rangle_{\beta}=\partial_0\partial'_0\langle0| A_i(x(\tau))A_j(x(\tau'))|0\rangle_{\beta}+\partial_i\partial'_j\langle0| A_0(x(\tau))A_0(x(\tau'))|0\rangle_{\beta},
\end{equation}
where $\partial'$ denotes the differentiation with respect to $x'$.

Now applying the trajectory of the atom at rest
\begin{equation}\label{a3}
 t(\tau)=\tau,\;x(\tau)=x_0,\;y(\tau)=y_0,\;z(\tau)=z_0,
\end{equation}
we find the field two point function can be evaluated to get
\begin{eqnarray}
\langle0|E^{f}_i(x(\tau))E^{f}_j(x(\tau'))|0\rangle_{\beta}=\frac{1}{\pi^{2}}\sum^{\infty}_{k=-\infty}\bigg[\frac{\delta_{ij}}{(u+ik\beta-i\varepsilon)^{4}}
-\frac{(\delta_{ij}-2n_in_j)(u+ik\beta)^{2}+4z^{2}_0}{[(u+ik\beta-i\varepsilon)^{2}-4z^{2}_0]^{3}}\bigg],\nonumber\\
\end{eqnarray}
with $u=\tau-\tau'$. Then the functions $\mathcal{G}^{+}(\omega_{0})$, $\mathcal{G}^{-}(-\omega_{0})$ can be calculated as
\begin{eqnarray}\label{ar}
\mathcal{G}^{+}(\omega_{0})&=&e^{2}\langle +|r_i(0)|-\rangle\langle -|r_j(0)|+\rangle \int^{\infty}_{-\infty} d u e^{i\omega_0 u}\langle0| E^{f}_i(x(\tau))E^{f}_j(x(\tau'))|0\rangle_{\beta}\nonumber\\
&=&\frac{e^{2}\omega^{3}_0}{3\pi}|\langle +|r_i(0)|-\rangle|^{2} (1-f_{i}(\omega_0,z_0))(1+\frac{1}{e^{\beta\omega_0}-1}),\nonumber\\
\mathcal{G}^{-}(-\omega_{0})&=&e^{2}\langle -|r_i(0)|+\rangle\langle +|r_j(0)|-\rangle \int^{\infty}_{-\infty} d u e^{-i\omega_0 u}\langle0| E^{f}_i(x(\tau))E^{f}_j(x(\tau'))|0\rangle_{\beta}\nonumber\\
&=&\frac{e^{2}\omega^{3}_0}{3\pi}|\langle +|r_i(0)|-\rangle|^{2}(1-f_{i}(\omega_0,z_0))\frac{1}{e^{\beta\omega_0}-1},
\end{eqnarray}
where $f_{x}(w_0,z_0)=f_{y}(\omega_0,z_0)=\frac{3}{16\omega^{3}_0z^{3}_0}\big[2\omega_0z_0\cos(2\omega_0z_0)+(4\omega^{2}_0z^{2}_0-1)\sin(2\omega_0z_0)\big]$ and $f_{z}(w_0,z_0)=\frac{3}{8\omega^{3}_0z^{3}_0}\big[2\omega_0z_0\cos(2\omega_0z_0)-\sin(2\omega_0z_0)\big]$ are oscillating functions of
distance $z_{0}$ with a position-dependent amplitude. It can be examined that when $z_0\rightarrow \infty$, $f_i(\omega_0,z_0)\rightarrow 0$ as expected and when $z_0\rightarrow 0$, $f_x(\omega_0,z_0)=f_y(\omega_0,z_0)=-f_z(\omega_0,z_0)=1$.
The term $1/(e^{\beta\omega_0}-1)$ gives thermally induced corrections to the Einstein $A$ coefficients which respectively are
\begin{equation}\label{e1}
  A_\downarrow = \frac{e^{2}\omega^{3}_0}{3\pi}|\langle +|r_i(0)|-\rangle|^{2} (1-f_{i}(\omega_0,z_0)),\; A_\uparrow = 0.
\end{equation}

Finally, as can be seen from (\ref{10}),  the total rate of change of the atomic energy becomes now
\begin{eqnarray}\label{e2}
 \bigg\langle\frac{dH_A(\tau)}{d\tau}\bigg\rangle_{tot}=\Bigg\{
         \begin{array}{r}
 -\omega_0\gamma_0\alpha_i(1-f_{i}(\omega_0,z_0))(1+\frac{1}{e^{\beta\omega_0}-1}), \; \;excited\; state,\\
   \omega_0\gamma_0\alpha_i(1-f_{i}(\omega_0,z_0))\frac{1}{e^{\beta\omega_0}-1}, \;\;ground\; state.\\
          \end{array}
\end{eqnarray}
Here $\gamma_0=e^{2}\omega^{3}_0|\langle +|\textbf{r}(0)|-\rangle|^{2}/{3\pi}$ denotes the spontaneous emission rate in vacuum
and $\alpha_i=|\langle +|r_i(0)|-\rangle|^{2}/|\langle +|\textbf{r}(0)|-\rangle|^{2}$ represents the relative polarizability and they
satisfy $\sum_i\alpha_i=1$. It can be found that in the case that the atom is placed near the boundary, for an isotropic polarization that is $\alpha_x=\alpha_y=\alpha_z=1/3$,
the total rate of change of the atomic energy is two thirds of the no-boundary case; for the polarization is along the $z$-axis that is $\alpha_x=\alpha_y=0$,
the total rate of change of the atomic energy becomes double of that without the boundary; for the polarization is in the $xy$
plane that is $\alpha_z=0$, the total rate of change of the atomic energy becomes zero. So atom with different polarizations will contribute differently to the spontaneous processes.

\section{The case of an accelerated atom near a reflecting boundary}
\label{sec:example2}
Let us now consider the case of the atom accelerating in the $x$-direction near a conducting plate located at $z=0$.
The trajectory of atom is described by
\begin{equation}\label{111}
t(\tau)=\frac{1}{a}\sinh a \tau,\;x(\tau)=\frac{1}{a}\cosh a \tau,\;y(\tau)=y_{0},\;z(\tau)=z_{0}.
\end{equation}
In such a case, the two point function of the electric
field four potential changes to
\begin{eqnarray}
  \langle0| A^{\mu}(x)A^{\nu}(x')|0\rangle&=&\frac{1}{4\pi^{2}}\bigg[\frac{\eta^{\mu\nu}}{(t-t'-i\varepsilon)^{2}-(x-x')^{2}-(y-y')^{2}-(z-z')^{2}}\nonumber\\
  &&-\frac{\eta^{\mu\nu}+2n^{\mu}n^{\nu}}{(t-t'-i\varepsilon)^{2}-(x-x')^{2}-(y-y')^{2}-(z+z')^{2}}\bigg].
\end{eqnarray}
From (\ref{a2}), the electric field correlation function can then be evaluated to get
 \begin{eqnarray}\label{222}
 \langle0|E^{f}_x(x)E^{f}_x(x')|0\rangle&=&\frac{1}{\pi^{2}}\bigg[\frac{-(t-t')^{2}+(x-x')^{2}-(y-y')^{2}-(z-z')^{2}}
 {[(t-t'-i\varepsilon)^{2}-(x-x')^{2}-(y-y')^{2}-(z-z')^{2}]^{3}}\nonumber\\
  &&-\frac{-(t-t')^{2}+(x-x')^{2}-(y-y')^{2}-(z+z')^{2}}
 {[(t-t'-i\varepsilon)^{2}-(x-x')^{2}-(y-y')^{2}-(z+z')^{2}]^{3}}\bigg].
 \end{eqnarray}
Note that, for simplicity, here we only consider the case that the polarization is along the $x$-axis. Now substituting
(\ref{111}) in (\ref{222}), we have
\begin{equation}
\langle0|E^{f}_x(x(\tau))E^{f}_x(x(\tau'))|0\rangle=\frac{a^{4}}{16\pi^{2}}\bigg(\frac{1}{\sinh^{4}\frac{a}{2}(u-i\varepsilon)}
+\frac{a^{2}z^{2}_{0}+\sinh^{2}\frac{a}{2}u}{[a^{2}z^{2}_{0}-\sinh^{2}\frac{a}{2}(u-i\varepsilon)]^{3}}\bigg),
\end{equation}
with $u=\tau-\tau'$.
Then we can obtain
\begin{eqnarray}\label{ccc}
\mathcal{G}^{+}(\omega_{0})=A_\downarrow&=&e^{2}|\langle +|r_x(0)|-\rangle|^{2} \int^{\infty}_{-\infty} d u e^{i\omega_0 u}\langle0| E^{f}_x(x(\tau))E^{f}_x(x(\tau'))|0\rangle\nonumber\\
&=&\gamma_{0}\alpha_{x} (1+\frac{a^{2}}{\omega^{2}_0}-f_{x}(\omega_0,z_0,a))(1+\frac{1}{e^{2\pi\omega_0/a}-1}),\nonumber\\
\mathcal{G}^{-}(-\omega_{0})=A_\uparrow&=&e^{2}|\langle +|r_x(0)|-\rangle|^{2} \int^{\infty}_{-\infty} d u e^{-i\omega_0 u}\langle0| E^{f}_x(x(\tau))E^{f}_x(x(\tau'))|0\rangle\nonumber\\
&=&\gamma_{0}\alpha_{x}(1+\frac{a^{2}}{\omega^{2}_0}-f_{x}(\omega_0,z_0,a))\frac{1}{e^{2\pi\omega_0/a}-1},
\end{eqnarray}
where $f_{x}(\omega_0,z_0,a)=\frac{3}{16\omega^{3}_0z^{3}_0}\big[\frac{4\omega^{2}_0z^{2}_0(1+a^{2}z^{2}_0)-2a^{2}z^{2}_0(1+2a^{2}z^{2}_0)-1}
{(1+a^{2}z^{2}_0)^{5/2}}\sin(\frac{2\omega_0\sinh^{-1}(az_0)}{a})
+\frac{2\omega_0z_0(1+4a^{2}z^{2}_0)}{(1+a^{2}z^{2}_0)^{2}}\cos(\frac{2\omega_0\sinh^{-1}(az_0)}{a}) \big]$ is a oscillating function induced by
the presence of the boundary. It can be verified that when $a\rightarrow 0$, $f_{x}(\omega_0,z_0,a)\rightarrow f_{x}(\omega_0,z_0)$.
Now inserting (\ref{ccc}) into (\ref{10}), we get the total rate of change of the atomic energy
\begin{eqnarray}\label{hhh}
 \bigg\langle\frac{dH_A(\tau)}{d\tau}\bigg\rangle_{tot}=\Bigg\{
         \begin{array}{r}
 -\omega_0\gamma_{0}\alpha_{x} (1+\frac{a^{2}}{\omega^{2}_0}-f_{x}(\omega_0,z_0,a))(1+\frac{1}{e^{2\pi\omega_0/a}-1}), \; \;excited\; state,\\
   \omega_0\gamma_{0}\alpha_{x}(1+\frac{a^{2}}{\omega^{2}_0}-f_{x}(\omega_0,z_0,a))\frac{1}{e^{2\pi\omega_0/a}-1}, \;\;ground\; state.\\
          \end{array}
\end{eqnarray}
Comparing (\ref{hhh}) with (\ref{e2}) in the case that the polarization is along the $x$-axis, it can be found that they are not equal to each other, no matter the plate exists or not. So the equivalence between uniform acceleration and thermal bath is not existed in electromagnetic field case, which is in contrast to the scalar field case, as has been pointed out in \cite{zzzz}.

\section{Conclusions}
\label{sec:1}
In summary, for a two-level atom in a bath of fluctuating electromagnetic field, we have given a new representation for the
rate of variation of the mean atomic energy as well as the contributions of vacuum fluctuations and radiation reaction.
In the case of an atom on arbitrary stationary trajectory, our results show that the rate of excited in ground
state and the rate of de-excited in the excited state by vacuum fluctuations are all equal to
$(\mathcal{G}^{+}(\omega_{0})+ \mathcal{G}^{-}(-\omega_{0}))/2$. It is de-excited by radiation reaction with the rate $(\mathcal{G}^{+}(\omega_{0})- \mathcal{G}^{-}(-\omega_{0}))/2$, no matter
which state the atom is initially in. For any initial excited state, the rate of change of atomic
energy is $-\omega_0\mathcal{G}^{+}(\omega_{0})$. In the ground state, it is $\omega_0\mathcal{G}^{-}(-\omega_{0})$. It is worthwhile to note that for the scalar field case, we can get a similar formalism.

Besides, we deduce the general expressions of the Einstein spontaneous emission and excitation coefficients from this new formalism.
We also give two cases to show the strength of our formalism.

\begin{acknowledgments}
We would like to thank Y. Jin for a valuable discussion.
\end{acknowledgments}

\end{document}